\documentclass[preprintnumbers,showpacs,aps,amsmath,amssymb,superscriptaddress,nofootinbib,twocolumn,10pt]{revtex4-1}
\usepackage{amssymb,amsmath}
\usepackage{hyperref}
\usepackage{color}
\usepackage{graphicx}
\usepackage{ulem}
\usepackage{array} 
\usepackage{epstopdf}

\DeclareGraphicsExtensions{.pdf,.eps,.png,.jpg,.mps}

\def\beq{\begin{equation}}
\def\eeq{\end{equation}}
\newcommand{\elE}{\mathcal{E}}
\newcommand{\elK}{\mathcal{K}}

\newcommand{\Exp}{\mathop{\mathrm{Exp}}}

\renewcommand\Im{\operatorname{Im}}
\DeclareMathOperator{\sech}{sech}

\begin{abstract}
It is now well established, both theoretically and experimentally, that very small changes in the size of isolated nanograins lead to substantial non-monotonic variations, and sometimes enhancement, of the mean-field spectroscopic gap of conventional superconductors. A natural question to ask, of broad relevance for the theory and applications of superconductivity, is whether these size effects can also enhance the critical temperature of a bulk granular material composed of such nanograins. Here we answer this question affirmatively. We combine mean-field, semiclassical and percolation techniques to show that engineered nanoscale granularity in conventional superconductors can enhance the critical temperature by up to a few times compared to the non-granular bulk limit. This prediction is valid for three dimensional and also quasi-two dimensional samples, provided the thickness is much larger than the grain size. Our model  
 takes into account an experimentally realistic distribution of grain sizes in the array, charging effects, tunneling by quasiparticles and limitations related to the proliferation of thermal fluctuations for sufficiently small grains. %For small normal resistances we find the transition is percolation driven. Whereas at larger resistances the transition occurs above the percolation threshold due to phase fluctuations. 
\end{abstract}

\begin{document}
\title{Strong enhancement of bulk superconductivity by engineered nano-granularity}
\author{J. Mayoh}
\author{A.M. Garc\'ia-Garc\'ia}
%\affiliation{CFIF, Instituto Superior T{\'e}cnico,
%Universidade de Lisboa, Av. Rovisco Pais, 1049-001 Lisboa, Portugal}
\affiliation{Cavendish Laboratory, University of Cambridge, JJ Thomson Av., Cambridge, CB3 0HE, UK}
\date{\today}
\maketitle
The quest for higher critical temperatures, $T_c$, is one of the main driving forces in the field of superconductivity. 
Tuning the number of charge carriers\cite{Caviglia2008}, increasing pressure\cite{Mizuguchi2008} or disorder\cite{Feigelman2007}, applying microwave radiation\cite{Kommers1977} or a high pulse field\cite{Fausti2011} and exploiting finite size effects\cite{Blatt1963,parmenter1968} are only a few of the mechanisms proposed to increase the $T_c$ of superconductors. The last of these, first proposed theoretically 
in the sixties\cite{Parmenter1968a,Blatt1963,*Thompson1963}, predicts more robust superconductivity in nanostructures by tuning the Fermi energy to a region of anomalously large spectral density. The effect is especially strong in symmetric nano-grains with level degeneracies, usually referred to as shells. Recent mean-field numerical\cite{Shanenko2006,*Shanenko2007,*Croitoru2007} and analytical\cite{Garcia-Garcia2011} results have confirmed that, for typical grain radii $R \sim 10$nm, enhancement of $T_c$ by size effects is still substantial. A mean-field approach is justified in this region since the mean level spacing, $\delta$, is much smaller than the bulk superconducting gap $\Delta_0$\cite{Anderson1959,Ralph1995,Richardson1963}.
Recent experimental results in isolated Sn nanograins\cite{Bose2010} are fully consistent with theoretical predictions.\\
\begin{figure}
\includegraphics[width=0.4\textwidth]{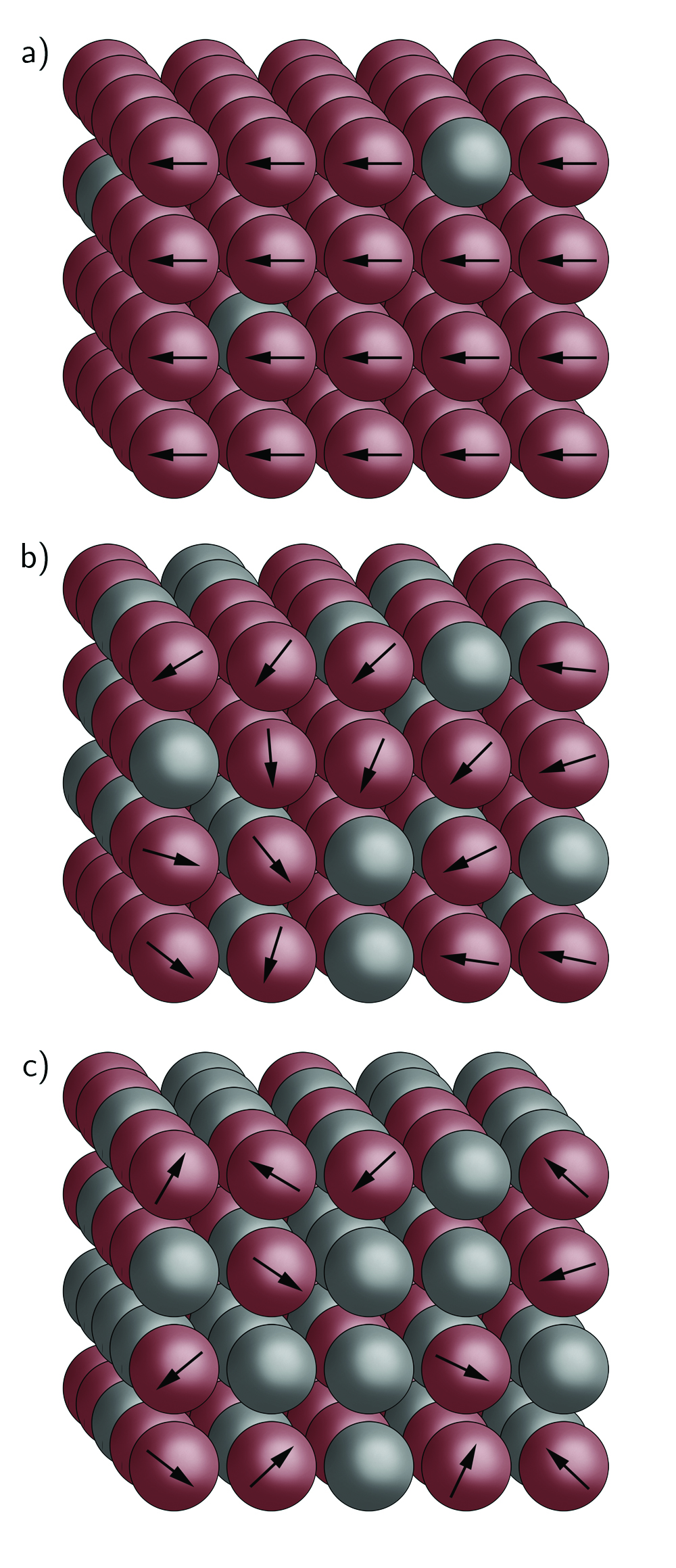}
 \caption{Sketch of the Josephson array of nanograins at three different temperatures: a) $T\ll T_c$, b) $T \lesssim T_c$, c) $T\gtrsim T_c$ where $T_c$ is the critical temperature of the grain in the bulk limit. The distribution of grain sizes is Gaussian with typical average $\sim 5$nm and variance $1$nm. Due to size effects the array is highly inhomogeneous as each grain has a different critical temperature. Red (gray) grains are (no longer) superconducting at that temperature. The phase of the superconducting gap is indicated by an arrow. The transition in the array occurs at a temperature for which either phase fluctuations induce the loss of phase coherence or there are not enough superconducting grains to form a cluster permeating the whole of the material. We have found a substantial enhancement of the array critical temperature for different average grain size and variance, packings, electron-phonon coupling and normal state resistance. For an optimal grain packing, see Fig. \ref{plot34}, the critical temperature of the array can be up to a few times higher than for a bulk non-granular material.} \label{fig0}
\end{figure}

However true superconductivity, characterized by phase-coherence and zero resistance, cannot exist in a single isolated grain. The number of particles inside the grain is fixed and hence the phase of the order parameter is delocalized. The situation could in principle be different in a Josephson array composed of such nanograins where inter-grain coupling can lead to bulk phase coherence. Due to size effects in the single grains a higher critical temperature\cite{parmenter1968} than in a bulk non-granular sample \cite{Fazio2001,Beloborodov2007} may be possible. Some experiments \cite{Abeles1966,Deutscher1973a,*Deutscher1973,*Shapira1982} indeed found enhancement of $T_c^{\text{Array}}$ in Al and other granular metallic superconductors \cite{Abeles1966,li2008} but no enhancement was observed in Sn\cite{Bose2005} or in samples where granularity was suppressed\cite{Orr1985,*Jaeger1986}. We note, see sketch in Fig. \ref{fig0}, that such arrays are intrinsically inhomogeneous as the critical temperature of neighboring grains can be very different. \\

With some exceptions \cite{Bianconi2012,*Bianconi2013} not much is known about the physical properties of inhomogeneous arrays. Moreover, size effects in single grains are obviously weakened by inter-grain coupling so it is not clear at all, a priori, whether nanogranularity can enhance superconductivity. Previous claims in the literature of orders of magnitude enhancement of the critical temperature  \cite{Ovchinnikov2012,*kresin2006,*Ovchinnikov2005a} do not take into account these features, namely, inhomogeneity of nanograins arrays and the weakening of size effects by inter-grain coupling, so their results are unrealistic.\\

Here we tackle this problem by putting forward a realistic model of a Josephson array of clean, superconducting nanograins. Explicit analytical results are obtained by combining mean-field, semiclassical and percolation techniques. The main goal of the letter is to clarify whether it is feasible to enhance bulk superconductivity by size effects in single nanograins and then to explore the set of realistic parameters that lead to the highest increase of the critical temperature of the array. Our results pave the way for the design of novel nano-engineered superconductors with tunable properties. \\
%with sizes extracted from a known random distribution. % with typical average $\sim 5$nm and variance $\sim 1$nm. 
%In Ref.\cite{Deutscher1973a} the distribution of grain sizes $\sim 5$nm was relatively narrow and it was also possible to tune the %resistance of the normal state. 

The formalism we use is applicable to nano-grains of any shape but we focus on 
spheres with negligible disorder since this is, not only, the geometry that leads to the strongest finite size effects but also the easiest one to fabricate experimentally\cite{Deutscher1973a,Bose2010}. We focus on a three dimensional array as global phase-coherence is easier to achieve in higher dimensionalities and a mean-field approach is more accurate. However, our results are also valid in quasi two dimensional geometries relevant for experiments, provided the thickness is much larger than the typical grain size. Indeed this is the case in most experiments\cite{Deutscher1973a,Bose2005,li2008}. The grain size in any realistic array \cite{Bose2010,Eley2011} is randomly distributed. Following the experimental results of Ref. \cite{Deutscher1973a} we employ a Gaussian distribution with a mean and variance of about $\sim 5$nm and $ 1$nm respectively. We stress that this implies some grains have a $T_c$ that is higher than the bulk material, $T_{c0}$, whilst for others it is lower.\\ 

The theoretical analysis is divided in two parts. First, we compute the weakening of size effects in the mean-field critical temperature of a single grain caused by coupling it to its nearest neighbors. We employ the mean-field Bardeen-Cooper-Schrieffer (BCS) formalism and semi-classical techniques that are only applicable in the limit $k_FR\gg 1$ and $T_{c0} \gg \delta$ so quantum and thermal fluctuations are negligible; where $k_F$ is the Fermi-wave vector and $R$ is the radius of the grain. Typically this limit corresponds to $R\geq 5$nm although the exact value depends on the material. Moreover, inter-grain coupling smooths out the spectrum and consequently enlarges the range of applicability of mean-field theory techniques.\\ 

Second, we compute the critical temperature of the array, defined as the temperature at which global phase coherence is lost, as a function of different parameters: the normal state resistance, the electron-phonon coupling constant, the grain size and way the grains in the array are arranged. Two mechanisms can induce the transition: phase fluctuations or that the fraction of grains that are superconducting  is not sufficient to form a cluster permeating the whole sample. The former mechanism is modeled by a mean-field formalism for the phase dynamics that includes charging effects, quasiparticle tunneling and the usual Josephson coupling that depends on the superconducting gap computed previously. The latter by counting the number of superconducting grains, defined as those with a finite superconducting gap, at a given temperature and comparing to the known results from percolation theory. The physical critical temperature is the lower of the two.\\

The main conclusion of this analysis is that it is possible by nano-engineering to enhance the critical temperature of a conventional superconductor by up to a few times the bulk non-granular limit. The optimal setting is for weakly coupled materials, such as Al, grain sizes $\sim 5$nm and FCC packing of the grains in the array. These results offer a plausible explanation of the observed enhancement of superconductivity in some granular materials. However quantitative comparisons with experiments will require better control of grain positioning such as that offered by poly-lithographic techniques where the material is built up layer by layer and the superconducting islands can be placed in a pattern with varying inter-island spacing\cite{Eley2011}.
A convincing experimental confirmation of these results would be a key step in the development of engineered superconductivity with tunable properties. 
% which leads to the largest enhancement of $T_c^{\text{Array}}$. 
%Based on recent experimental results \cite{Bose2010} it is also plausible to assume that disorder %inside the grains is negligible and the dynamics inside each grain is ballistic. The effective %potential in the grains is simply an infinite well with the shape of the grain. 
We start with a detailed theoretical description of the coupling of a single nanograin to the rest of the array.
\section{Model of a single grain coupled to the nearest neighbors}
We model the coupling of a single grain to the rest of the array using semiclassical techniques and a mean-field formalism. The overall effect of the coupling is a smoothing of the density of states that suppresses finite size effects.
Superconductivity in each grain is described by the BCS\cite{Bardeen1957} Hamiltonian,
\begin{equation}
H=\sum_{n\,\sigma}\epsilon_n c^\dag_{n\sigma}c_{n\sigma}-\frac{\lambda}{\nu_{TF}(0)}\sum_{n,n'}I_{n,n'}c_{n\uparrow}^\dag c_{n\downarrow}^\dag c_{n'\uparrow}c_{n'\downarrow}
\end{equation}
where $c_{n\sigma}^\dag$ creates an electron of spin $\sigma$ in state $n$ with energy $\epsilon_n$, $\lambda$ is the dimensionless BCS coupling constant, $\nu_{TF}(0)$ is the bulk density of states at the Fermi energy, $\epsilon_F$, and $I_{n,n'}$ are the short range electron-electron interaction matrix elements. The second sum is taken over all of the states within the Debye energy, $\epsilon_D$, window around $\epsilon_F$. The superconducting gap $\Delta(R,T)$ is given by,
\begin{equation}
1=\frac{\lambda}{2}\int_{-\epsilon_D}^{\epsilon_D}\frac{1}{\sqrt{\epsilon'^2+\Delta^2}}\frac{\nu(\epsilon')}{\nu_{TF}(0)}\tanh\left(\frac{\beta\sqrt{\epsilon'^2+\Delta^2}}{2}\right)d\epsilon'
\label{GapEqn}
\end{equation}
where $\beta=1/k_BT$ and $\nu(\epsilon)=\sum_n\delta(\epsilon-\epsilon_n)$ is the exact single particle density of states. Here $\nu(\epsilon)$ is dependent on the size of the grain and is the parameter responsible for including size effects in the model.
For simplicity we assume $I_{n,n'}=1$ which underestimates the size effects on the gap and $T_c$.

The most important difference between an isolated grain and one coupled to an array is that in the latter quasi-particles can escape by tunneling. The grain is therefore open and its density of states is smoothed. This smoothing is modeled by expressing the density of states analytically as a sum over classical periodic orbits with a cut-off that depends on the probability of inter-grain tunneling. The latter is a function of the tunneling resistance of the junction $R_N$ and the number of nearest neighbor. Explicit expressions of these quantities for the case of spherical grains are given in the appendix. 

Explicit expressions for $T_c(R)$ and $\Delta(R,T=0)$ as a function of the grain radius $R$ are then obtained from Eq.(\ref{GapEqn}) by a power expansion in the small parameter $(k_FR)^{-1/2}$\cite{Garcia-Garcia2011},
% \begin{equation}
% k_BT_c(R) \approx \frac{2\epsilon_De^\gamma}{\pi}e^{-\frac{1}{\lambda(\bar g(0)+\delta g_{T}(0))}}
% \end{equation}
% where $\gamma= 0.5772$ is the Euler-Mascheroni constant, $\bar g(0) = 1 - \frac{3\pi}{4k_FR} \ldots$ and $\delta g_{T}(0)$ is given by Eq.(\ref{densi}) with $\omega(R_N,L_P) \to \omega(R_N,L_P)\int_0^{\epsilon_D}\frac{1}{\epsilon}\cos(\frac{\epsilon}{2\epsilon_F} k_FL_P)\tanh(\frac{\beta\epsilon}{2})d\epsilon/\ln(\frac{2e^\gamma\beta\epsilon_D}{\pi})$. 
The superconducting gap close to $T_c$ is given by,
\begin{equation}
\Delta(R,T) \approx 1.74\Delta(R,0)\left(1-\frac{T}{T_c(R)}\right)^\frac{1}{2}.
\end{equation}
These expressions will be the key building blocks to study the behavior of the array. For more details of these calculations we refer to the appendix.
\section{Model of the Josephson array based on coupled nano-grains}
We now turn to the theoretical description of an array composed of the spherical nanograins studied in the previous section. In order to mimic realistic experimental conditions\cite{Deutscher1973a,Bose2010} we consider a Gaussian distribution of grain sizes, $P(R)$, with mean $\bar R$ and standard deviation $\sigma$. As a consequence $T_c$ and the gap $\Delta$ are different in each grain. The fraction of grains in the normal metal phase increases as temperature increases. We choose a three dimensional array for which a Kosterlitz-Thouless transition\cite{Kosterlitz1973} is not favorable since, even close to the percolation threshold, the dimensionality of the percolating cluster is $D_f\sim 2.52 > 2$\cite{Ballesteros1999}. The grains that belong to the superconducting cluster are those that verify $T_c(R) \geq T_c^{\text{Array}}$, they have a distribution $P_{sc}(R)=\theta(T_c(R)-T)P(R)$. 

We aim to compute $T_c^{{\text{Array}}}$ as a function of $R_N$, $P(R),\lambda$ and the way the grains are packed in the array. There are two distinct ways to destroy global phase-coherence in the array. First, the array may reach its percolation threshold $p=p_c$ where $p$ is the fraction of grains in the superconducting phase, $p=\int_0^\infty P_{sc}(R)dR$, and $p_c$ is the percolation threshold of the array. Beyond the percolation threshold there may exist globally phase coherent clusters but these do not permeate the whole array. Second, global phase-coherence may be destroyed by phase-fluctuations. In the former the critical temperature is defined as the temperature for which the number of grains still superconducting, computed using the expressions obtained previously, form the critical percolating cluster. The calculation of the latter requires a more elaborate treatment. We start by considering the usual action, see appendix for a definition of the action, for this type of array \cite{Chakravarty1987} that includes charging effects, quasiparticle tunneling and the Josephson coupling, which is highly inhomogeneous as the value of the superconducting gap is different in each grain. 
%\begin{widetext}
%\beq\label{action}
%\begin{split}
%S=\frac{1}{2}\int_0^\beta d\tau\sum_i\frac{\dot{\phi_i^2}}{E_Q}-\frac{1}{2}\sum_{\langle ij\rangle}\int_0^\beta d\tau %J_{ij}\cos(2(\phi_i(\tau)-\phi_j(\tau)))+2\sum_{\langle ij\rangle}\int_0^\beta d\tau\int_0^\beta  d \tau' G_{ij}(\tau-\tau')\sin^2(\frac{1}%{4}(\delta\phi_{ij}(\tau)-\delta\phi_{ij}(\tau')))
%\end{split}
%\eeq
%\end{widetext}
%where $\phi_i$ is the phase in the grain $i$ and $\delta\phi_{ij}=\phi_i-\phi_j$. The first term is the charging energy for a grain with %capacitance $C$, $E_Q=\frac{4e^2}{C}$, the second term is the Josephson coupling\cite{Ambegaokar1963} $J_{ij}=\frac{\Delta_i\Delta_j}%{\beta}\frac{R_Q}{R_N}\sum_{l=-\infty}^{\infty}\frac{1}{\sqrt{((\frac{\pi(2l+1)}{\beta})^2+\Delta_i^2)%((\frac{\pi(2l+1)}{\beta})^2+\Delta_j^2)}}$, and the third term accounts for quasi-particle %tunneling\cite{Eckern1984}, $
%G_{ij}(\tau)=\frac{\hbar}{2\pi e^2R_N}\frac{\Delta_i\Delta_j}{\hbar^2}K_1\left(\frac{\Delta_i|\tau|}{\hbar}\right)K_1\left(\frac{\Delta_j|%\tau|}{\hbar}\right)
%$ where $\Delta_i$ is the gap in grain $i$ and $K_1$ is the modified Bessel function of order 1.

Here we only provide a broad overview of the calculation and refer to the appendix for further technical details. First we remove the position dependence of the Josephson coupling term by expressing it in terms of the mean gap $\bar\Delta_{ij}=\frac{\Delta_i+\Delta_j}{2}$ and the difference in the gaps $\Delta_{ij}'=\frac{|\Delta_i-\Delta_j|}{2}$ across the junction and expand in powers of $\Delta_{ij}'$. Then we approximate the superconducting cluster by a homogeneous array with $\bar\Delta_{ij}$ replaced by the mean value for the cluster,
$
\bar\Delta=\frac{1}{p^2}\int_0^\infty\int_0^\infty\frac{\Delta(R)+\Delta(R')}{2}P_{sc}(R)P_{sc}(R')dRdR'.
$
A similar procedure is applied to $\Delta'_{ij}$. This is a good approximation as close to $T_c^{\text{Array}}$ the distribution of $\bar\Delta_{ij}$ and $\Delta'_{ij}$ in the cluster will be narrow and sharply peaked around this value. We also introduce the mean number of superconducting neighbor grains in the percolating cluster, $\bar z=z p$. This value slightly underestimates the coordination number of the infinite cluster as this is the mean coordination number for the whole array including both finite clusters and the infinite cluster, however this discrepancy is small.

The resulting homogeneous action, already studied in the literature\cite{Panyukov1987}, can be tackled by standard mean-field techniques. 
%After a Hubbard-Stratonovich transformation in the Josephson term and integrating out $\phi_i(\tau)$ we find
%\beq
%\begin{split}
%\frac{S}{\hbar}=&\frac{\bar zJ}{2k_BT}\sum_{{\bf q}\;\omega}\left(1-\frac{\bar z J}{2}\int_0^{\beta\hbar}\frac{d\tau}{\hbar} X_{ii}%(\tau)\right)|\psi({\bf q,\omega})|^2\\ &+\frac{\zeta}{4}\sum_{{\bf q}\;\omega}|\psi({\bf q}, \omega)|^4
%\end{split}
%\eeq
%where $\psi$ is the Hubbard-Stratonovich field, $X_{ii}$ is the correlation function,
%$
%\ln{X_{ii}(\tau)}=-\frac{\tilde E_Q\tau}{2\hbar},
%$
% $\zeta$ is a numerical constant, 
The critical temperature of the array, due to phase-fluctuations, is obtained by finding the solution to,
\begin{equation}\label{CritCond}
1=\frac{\tilde E_Q}{ \bar z J}+e^{-\beta{\tilde E_Q}/2}
\end{equation}
 where $\tilde E_Q=(\frac{1}{E_Q}+\frac{\eta}{E_Q^*})^{-1}$,  $E_Q^*=\frac{124e^2\bar\Delta R_N}{3\pi\hbar}$, $J=\frac{\bar\Delta R_Q}{2R_N}\tanh(\frac{\beta\bar\Delta}{2})-\Lambda$, see appendix for full definition of $\Lambda$ . The values of $z, \eta, p_c$ for several geometries are summarized in Table \ref{Table}. Having determined critical conditions for the breaking of global phase-coherence due to percolation and phase-fluctuations we then define the critical temperature of the array, $T_c^{\text{Array}}$, to be the lower of the two critical temperatures. We note that we are assuming that inter-grain coupling is constant however the distance between grains in realistic arrays is rather random but with a well defined average and small variance. We believe that this is a fair approximation as random coupling only affects the percolation transition through the weakening of size effects in single grains which is a small correction. Regarding phase fluctuation we expect that random couplings will lower the critical temperature of the array. However, in the range of parameters we study this should not affect the maximum of the critical temperature which occur far from the quantum resistance where phase fluctuations are important.      
 \begin{figure}[b]
    \includegraphics[trim={5.12cm 19.05cm 10.3cm 4.55cm},clip,width=0.45\textwidth]{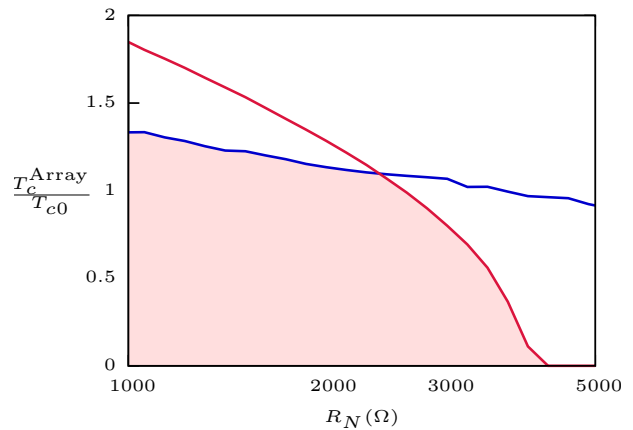}
 \caption{The critical temperature of the array in units of the bulk critical temperature of the material against $R_N$ for a cubic array with $\epsilon_F=10.2$eV, $\epsilon_D=9.5$meV, $\bar R =5$nm, $\sigma = 1.0$nm and $\lambda=0.3$. The blue line shows the critical temperature due to percolation given by finding the temperature  at which $p=p_c$. The red line shows the critical temperature due to phase-fluctuations found by solving Eq.\ref{CritCond}. Close to $2.5k\Omega$ these two lines cross meaning the transition that breaks global-phase coherence goes from being percolation to phase-fluctuation driven. The shaded region shows the range of parameters for which the array will be globally superconducting. This cross-over corresponds to the sharp tail seen at large resistance in the following figures.} \label{fig01}
\end{figure}
In the next section we explore the range of parameters that give the greatest enhancement of $T_{c}^{\text{Array}}$.
\begin{figure}[t]
\centering
\includegraphics[trim={5.12cm 19.05cm 10.55cm 4.65cm},clip, width=0.45\textwidth]{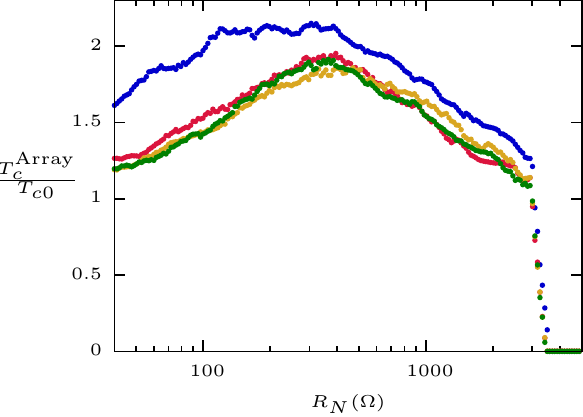}
\includegraphics[trim={5.12cm 19.05cm 10.55cm 4.4cm},clip, width=0.45\textwidth]{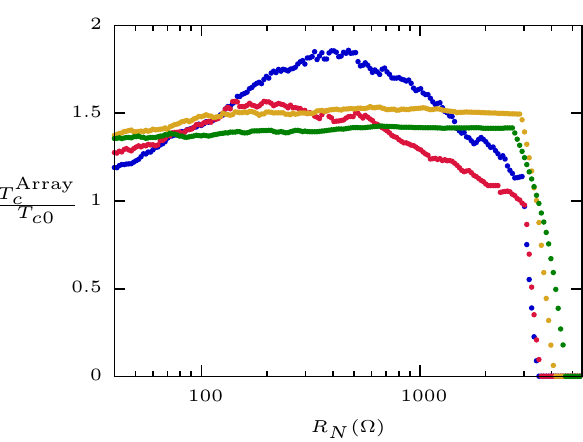}
\caption{$T_{c}^{\text{Array}}$ in units of the bulk material critical temperature, against $R_N$ for a cubic array with $\epsilon_F=10.2$eV, $\epsilon_D=9.5$meV, $\lambda=0.25$.\\
 {\it Top}: Gaussian distribution of sizes with mean $\bar R = 5$nm, and variance $\sigma=0.1$nm (blue, solid),$0.6$nm (red), $1.0$nm (yellow) and $1.4$nm (green). Results are weakly dependent on $\sigma$ (for $\sigma > 1 \AA$) as the typical scale for which shell effects in the size distribution are not randomized is much smaller.\\
{\it Bottom}: $\sigma = 1.0$nm and $\bar R =5$nm(blue), $7$nm(red), $11$nm(yellow), $17$nm(green). $T_c^{\text{Array}}$ becomes independent of $R_N$, due to the decreasing importance of quasi-particle tunnelling, and then gradually decreases for increasing $\bar R$ due to the weakening of shell effects}\label{plot12}
\end{figure}
% \begin{figure}[t]
% \centering
% \begin{tabular}{cc}
% \includegraphics[width=0.22\textwidth]{dR.eps}&
% \includegraphics[width=0.22\textwidth]{Rbar.eps}
% \end{tabular}
%  \caption{$T_{c}^{\text{Array}}$ in units of the bulk material critical temperature, against $R_N$ for a cubic array with $\epsilon_F=10.2$eV, $\epsilon_D=9.5$meV, $\lambda=0.25$.
%  {\it Left}: Gaussian distribution of sizes with mean $\bar R = 5$nm, and variance $\sigma=0.1$nm (blue, solid),$0.6$nm (red, dotted), $1.0$nm (yellow, dashed) and $1.4$nm (green, dot-dashed). Results are weakly dependent on $\sigma$ (for $\sigma > 1 \AA$) as the typical scale for which shell effects in the size distribution are not randomized is much smaller.
% {\it Right}: $\sigma = 1.0$nm and $\bar R =5$nm(blue, solid), $7$nm (red, dotted), $11$nm(yellow, dashed), $17$nm(green, dot-dashed). $T_c^{\text{Array}}$ becomes independent of $R_N$, due to the decreasing importance of quasi-particle tunneling, and then gradually decreases for increasing $\bar R$ due to the weakening of shell effects} \label{plot12}
% \end{figure}
\section{Results}
We compute $T_c^{\text{Array}}$  assuming the the grain size distribution is a Gaussian,
$P(R)= \frac{1}{\sqrt{2\pi}\sigma}e^{-\frac{(R-\bar R)^2}{2\sigma^2}}$. This choice with $\sigma \sim 1$nm and $\bar R \sim 5$nm is a good approximation to the experimental distribution\cite{Deutscher1973}. Throughout the calculation we use $C = 4\pi\epsilon_0R\sim0.5aF$. However this capacitance is typically strongly renormalized, see Eq. \ref{CritCond}, by the quasiparticle tunneling so its value does not influence our results. As mentioned above the physical $T_c^{\text{Array}}$ is the minimum of the critical temperature computed from Eq.(\ref{CritCond}) and the critical temperature at which the array reaches its percolation threshold $p=p_c$. In Fig. \ref{fig01} we depict  both critical temperatures for typical values of the grain size and tunneling resistance. A percolation driven transition is observed for $R_N \lesssim R_Q$. For larger $R_N$ phase fluctuations, induced by $\tilde E_Q$, break long range order at temperatures below the percolation transition. %This results is robust to small changes of the parameters though we cannot rule out that in realistic granular materials the effective capacitance is not given by the classical expression above.

We first investigate the dependence of $T_c^{\text{Array}}$ on the width of the distribution $\sigma$. For experimentally realistic values, see Fig. \ref{plot12}, the results depend very weakly on $\sigma$. This is expected as oscillations in the order parameter due to shell effects take place on a much smaller length scale $\sim 1\AA$. Indeed when we tune $\sigma$ to this range we start to see substantial deviations depending on whether shell effects enhance or suppress $T_c(R)$ for $R = \bar R$. However, it is not realistic to expect such a narrow distribution to be experimentally feasible in the near future. %For $R_N \gtrsim R_Q$ the $T_c^{\text{Array}}$ drops sharply in all cases as the renormalized charging energy, $\tilde E_Q$, becomes important and the transition is driven by phase fluctuations.

Consider next the behavior of the array as $R_N$ increases, we observe a peak $\sim 500 \Omega$ indicating there is an optimal coupling strength for the array. In general, we expect an increase in $T_c^{\text{Array}}$ as $R_N$ increases due to the decreasing strength of inter-grain coupling. This makes the shell effects within each grain larger meaning some grains now have a significantly enhanced $T_c$. However, for sufficiently large $R_N \lesssim  R_Q$ there is very little smoothing of the spectral density in single grains. This results in a lower $T_c^{\text{Array}}$ as the fraction of grains with an enhanced $T_c$ is not sufficient to form a percolating cluster. This is the reason for the peak observed at intermediate $R_N$.

We then move to the dependence of $T_c^{\text{Array}}$ on the mean grain size $\bar R$. For large $\bar R$ results should be less dependent on $R_N$ as in this case the width of the peaks in the density of states is not controlled by $R_N$ but rather by the coherence length $\xi=\hbar v_F/\Delta_0$. Finite size effects diminish as $\bar R$ increases %(up to $\sim 60$nm)
which results, see Fig. \ref{plot12}, in a smaller enhancement of $T_c^{\text{Array}}$ as the $T_c$ of the single grains is not increased as much. We restrict ourselves to $\bar R > 5$nm so that thermal and quantum fluctuations, that break the mean-field theory approach, are unimportant.  
\begin{table}
\begin{center}
    \begin{tabular}{c @{\hskip 10pt} c@{\hskip 10pt} c@{\hskip 10pt} c}
    Packing Geometry& z & $\eta$ & $p_c$ \\ \hline
    Simple Cubic & 6 & 5.9 & 0.3116  \cite{Grassberger1992} \\ 
    Body Centered Cubic(BCC) & 8 & 5.4 & 0.2460  \cite{Bradley1991} \\
    Face Centered Cubic(FCC) & 12 & 5.1 & 0.1992  \cite{Lorenz1998} \\
    \end{tabular}
    \caption{Intrinsic properties of the three most common packing geometries. From left to right: packing, coordination number, integration constant in $\tilde E_Q$ (see below Eq (\ref{CritCond})) and the site percolation threshold.}\label{Table}
\end{center}
\end{table} 
 Significantly we observe that, for a broad range of $\bar R$, $T_c^{\text{Array}}$ is well above that of a non-granular bulk material. This is a quite general result that only requires a three dimensional array is inhomogeneous with a distribution of $T_c(R)$ around the bulk value $T_{c0}$. 

\begin{figure}[t]
\centering
\includegraphics[trim={5.12cm 19.05cm 10.55cm 4.65cm},clip, width=0.45\textwidth]{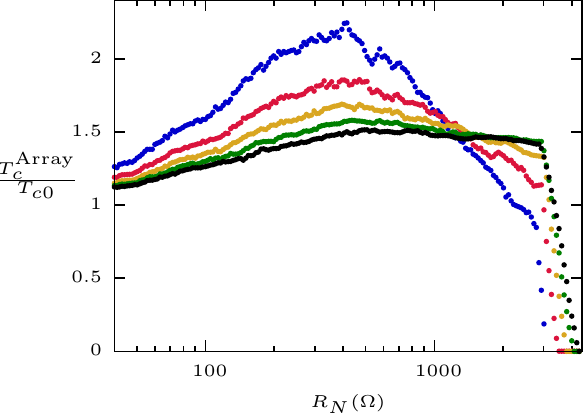}
\includegraphics[trim={5.12cm 19.05cm 10.55cm 4.4cm},clip, width=0.45\textwidth]{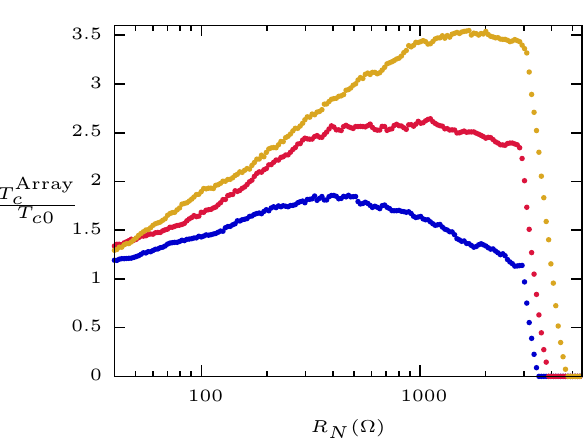}
\caption{$T_c^{\text{Array}}$ in units of the bulk material critical temperature against $R_N$ with $\epsilon_F=10.2$eV, $\epsilon_D=9.5$meV, $\bar R =5$nm, $\sigma = 1.0$nm.\\
{\it Top}: A cubic array with $\lambda=0.2$(blue), $0.25$(red), $0.3$(yellow), $0.35$(green), $0.40$(black). Increasing $\lambda$ suppress size effects resulting in a behavior which is closer to the bulk. \\
{\it Bottom}: $\lambda=0.25$ for a cubic (blue), BCC (red) and FCC (yellow) array. A smaller $p_c$ substantially enhances $T_c^{\text{Array}}$ by allowing the removal of more grains from the superconducting cluster so that the remaining ones have a higher $T_c$}\label{plot34}
\end{figure}

The value of the BCS coupling constant, $\lambda$, also plays an important role in $T_c^{\text{Array}}$. The larger $\lambda$, the less important size effects are. This follows from the fact that the coherence length $\xi$ decreases as  $\lambda$ increases thus making the material more bulk like. This prevents the employment of our BCS theory based approach in the study of cuprates and other strongly coupled superconductors. The results depicted in Fig. \ref{plot34} fully confirm this picture. 

Strikingly the array geometry, the way spheres are packed in the array, has a substantial effect on $T_c^{\text{Array}}$.
Settings which decrease the percolation threshold allow the array to remain globally phase-coherent with fewer superconducting grains. This results in a much higher  $T_c^{\text{Array}}$, see Fig. \ref{plot34}. The peaks for intermediate $R_N$ also moves to larger values of the resistance since the coupling between grains becomes stronger with increasing $z$.
 
% \begin{figure}[h]
% \centering
% \begin{tabular}{cc}
% \includegraphics[width=0.22\textwidth]{lambda.eps}&
%  \includegraphics[width=0.22\textwidth]{geom.eps}
% \end{tabular}
%  \caption{$T_c^{\text{Array}}$ in units of the bulk material critical temperature against $R_N$ with $\epsilon_F=10.2$eV, $\epsilon_D=9.5$meV, $\bar R =5$nm, $\sigma = 1.0$nm.
% {\it Left}: A cubic array with $\lambda=0.2$(blue, solid), $0.25$(red, dotted), $0.3$(yellow, dashed), $0.35$(green, dot-dashed), $0.40$(Black, double-dot-dashed). Increasing $\lambda$ suppress size effects resulting in a behavior which is closer to the bulk. 
% {\it Right}: $\lambda=0.25$ for a cubic (blue, solid), BCC (red, dotted) and FCC (yellow, dashed) array. A smaller $p_c$ substantially enhances $T_c^{\text{Array}}$ by allowing the removal of more grains from the superconducting cluster so that the remaining ones have a higher $T_c$} \label{plot34}
% \end{figure}
\begin{figure}[h!]
    \includegraphics[trim={5.05cm 19.0cm 10.5cm 4.6cm},clip,width=0.45\textwidth]{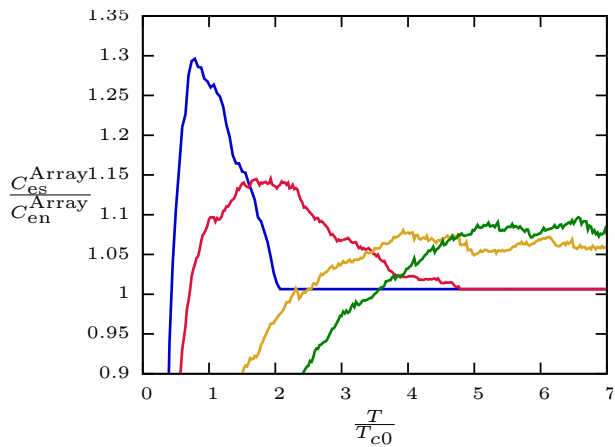}
 \caption{The superconducting electronic specific heat of the array in units of the normal electronic specific heat as a function of the temperature for  $\lambda=0.2$, $z=6$, $\bar R =5$nm, $\sigma=0.5$nm and different values of the tunneling resistance $R_N=50\Omega$(blue),$100\Omega$(red),$500\Omega$(yellow),$1000\Omega$(green). As the resistance increases the peak of the specific heat becomes broader. This is a signature of a percolation-driven transition.} \label{fig02}
\end{figure}
The outcome of this detailed analysis is that the maximum increase of $T_c^{\text{Array}}$, with respect to the bulk limit, is found in arrays of weakly coupled superconductor, $\lambda \ll 1$ with a mean grain size $\sim 5$nm, for intermediate resistances and for packings with a minimal percolation threshold. An interesting option to further increase $T_c^{\text{Array}}$ is to go beyond the minimum percolation threshold, corresponding to FCC, by employing two types of grains each one with a different average size. As discussed in Ref.\cite{Hudson2011} such an arrangement can be packed with a higher coordination number and hence a lower percolation threshold. Therefore a higher $T_c^{\text{Array}}$ is also expected.

We have also computed analytically the specific heat $C_\text{es}$ of the array from the entropy $S$, $C_\text{es} = -\beta\, \partial S /\partial \beta$ (see appendix for more details). This is an interesting observable as it is easy to measure experimentally and it can be used to detect a non bulk transition in which spatial inhomogeneities are important. A transition with a highly inhomogeneous order parameter is characterized by a broad peak around the critical temperature while for more homogeneous systems the transition is more bulk like with a shaper peak.   
As is shown in Fig. \ref{fig02} the specific heat becomes much broader than the bulk BCS prediction but enhancement of $T_c$ is still observed. The specific heat peak broadens as the tunneling resistance $R_N$ increases, in line with the experimental results of \cite{Worthington1978}. This prediction, relatively easy to test experimentally, is a clear signature that the enhancement of the critical temperature is not a bulk effect but rather it is related to the percolation of a critical superconducting cluster due size effects of the single nano-grains. 
Our plot of specific heat for an array of nanograins exhibits long tails extending to very high temperature. These tails are caused by a vanishingly small fraction of the grains which, in our model, are predicted to have a very high critical-temperature. We do not expect such grains to be realized experimentally. It is very likely the  anomalously large spectral density around the Fermi energy in these grains causes electronic or lattice instabilities in a perfectly spherical shape that reduce the critical temperature.  However the fraction of grains this applies to is extremely small so the results for the critical temperature of the array are not modified even if these grains are not taken into account.

In summary we have studied the properties of a large three dimensional array of superconducting spherical nano-grains. We have shown that superconductivity in the array is enhanced by the shell effects of the single grains. Our model includes a realistic distribution of grain sizes and tunneling to the nearest grains.  For $\lambda \sim 0.25$, FCC packings, $\bar R \sim 5$nm and $R_N \sim 1k \Omega$ we have observed that $T_c^{\text{Array}}$ can be more than three times higher than in the non-granular bulk limit. This values are in reasonable agreement with the experimental results in ref. \cite{Deutscher1973a,*Deutscher1973,*Shapira1982} where a peak critical temperature was observed $R_N\sim1k\Omega$. These results pave the way for technological applications that exploit size effects to engineer materials with more robust superconductivity. 

\vspace{0.45 cm}
{\bf Acknowledgments}\\
We thank Nimrod Bachar and Sangita Bose for fruitful discussions. JM acknowledges support from an EPSRC Ph.D. studentship. AMG acknowledges support from EPSRC, grant No. EP/I004637/1,  FCT, grant PTDC/FIS/111348/2009 and a Marie Curie International Reintegration Grant
PIRG07-GA-2010-268172\\
%%%%%%%%%%%%%%%%%%%%%%%%%%%%%%%%%%%
%%%%%%%%%%%%%%%%%%%%%%%%%%%%%%%%%%%%

%\documentclass[reprint]{revtex4-1}
%\usepackage{amssymb,amsmath}
%\usepackage{hyperref}
%\usepackage{color}
%\usepackage{graphicx}
%\usepackage{ulem}
%\newcommand{\Ci}{\mathop{\mathrm{Ci}}\nolimits}
%\newcommand{\sinc}{\mathop{\mathrm{sinc}}\nolimits}
%\def\beq{\begin{equation}}
%\def\eeq{\end{equation}}
%\newcommand{\elE}{\mathcal{E}}
%\newcommand{\elK}{\mathcal{K}}
%\newcommand{\calD}{\mathcal{D}}
%\newcommand{\arcsinh}{\mathop{\mathrm{ArcSinh}}}
%\newcommand{\Exp}{\mathop{\mathrm{Exp}}}
%\newcommand{\erfi}{\mathop{\mathrm{Erfi}}}
%\renewcommand\Im{\operatorname{Im}}
%\DeclareMathOperator{\sech}{sech}
%\newcommand{\C}{\color{red}}
% \setlength{\tabcolsep}{5pt}
%
%\begin{abstract}
%\end{abstract}

%\begin{document}
%\title{Strong enhancement of bulk superconductivity by engineered nano-granularity appendix}
%\author{J. Mayoh, A.M.Garc\'ia-Garc\'ia}
%\affiliation{Theory of Condensed Matter Group, University of Cambridge}
%\date{\today}

%\maketitle#
\appendix*
\section{Calculation details}
In this appendix we give some details of the calculations which are skimmed over in the main text. 

\subsection{Density of states in coupled Grains}
 We find this local density of states for a grain which is coupled to an array by extending the Gutzwiller trace formalism\cite{Brack1997}.

The density of states for isolated, spherical grains is given by,
\begin{equation}\label{DOS1}
\nu(\epsilon)=\nu_{TF}(\epsilon)(\bar{g}(\epsilon)+\delta g(\epsilon))
\end{equation}
where $\nu_{TF}(\epsilon)=\frac{2(kR)^3}{3\pi\epsilon}$, $\bar{g}(\epsilon)$ is the Weyl expansion for the density of states, here taken to have Dirichlet boundary conditions,
\begin{equation}\label{DOS2}
\bar{g}(\epsilon)=\left(1-\frac{3\pi}{4}\frac{1}{kR}+\frac{1}{(kR)^2}\right)
\end{equation}
and the oscillating contribution, $\delta g$, is given by the trace over periodic orbits of the Green's functions $\mathcal{G}({\bf r},{\bf r},E)$,
\begin{equation}\label{GF}
\delta g=-\frac{1}{\pi\nu_{TF}(\epsilon)}\Im\int \mathcal{G}({\bf r},{\bf r},E +i\varepsilon)d{\bf r}\,\,\,\,\,(\varepsilon>0)
\end{equation}

This formalism can be extended to include coupling by modifying the Green's functions in Eq. (\ref{GF}) to allow for tunneling. In general one must also be careful to consider potential modification to the smooth terms $\nu_{TF}(\epsilon)$ and $\bar{g}(\epsilon)$ as when the barrier is removed completely the volume in these formulae has changed from that of the grain to the entire array and the boundary conditions likewise disappear changing the Weyl expansion. We neglect these modifications by considering only arrays which are weakly coupled where the boundary conditions are Dirichlet like and the effective volume is that of the grain. We will return later to consider the quantitative meaning of this limit and show this is in fact the experimentally interesting regime.

We consider a grain to be coupled to its nearest neighbors by a finite square potential barrier of height $V_0$ and width $a$. Within periodic orbit theory We are free to choose an orthonormal basis to treat the problem so we make the usual choice,
\begin{equation}
\psi(r)=
\begin{cases}
Ae^{ikr}+Be^{-ikr}& r<R \\
Fe^{\kappa r}+Ge^{-\kappa r}& R<r<R+a\\
Ce^{ikx}& R+a<r\\
\end{cases}
\end{equation}
where $\kappa=\sqrt{2m(V_0-E)}/\hbar$ and the prefactors A to F are constants set by the boundary conditions. Normalization is controlled through appropriate choice of $A$. In the weak coupling regime where $\kappa a > 1$ the wavefunctions can be normalized by assuming the wavefunction is entirely contained in the grain, neglecting the small leak into the barrier. In this limit $A=B= \frac{1}{\sqrt{V}}$.

The tunneling rate out of the grain is determined by Fermi's golden rule. As we are interested in the properties of the system in the temperature regime where many grains are normal and the remaining superconducting grains are close to their critical temperature we use the normal-normal tunneling form of Fermi's golden rule. For an electron moving from grain 1 to any of the $z$ neighboring grains,
\begin{equation}
I_{1\rightarrow z}=\frac{z}{eR_N}\int^\infty_{-\infty} f(E)(1-f(E)) d\,E
\end{equation}
where $R_N=(4\pi e^2|T|^2\nu(0)^2/\hbar)^{-1}$ is the normal state tunneling resistance of the junction and $f(E)$ is the Fermi-Dirac distribution. The tunneling rate per electron close to the Fermi-level then is given by,
\begin{equation}
\Gamma_T=\frac{z}{e^2R_N\nu(0)}
\end{equation}
To find the probability an electron tunnels during the time it takes to travel around a  periodic orbit $L_P$ we integrate this rate along the orbit path. Hence the probability  the electron is still inside the grain after a complete orbit is,
\begin{equation}
\int_0^{L_P}\frac{\Gamma_T}{v_F}\;dl= 1- \frac{4z L_P R_Q}{R_N\nu(0)v_Fh}\approx e^{-\frac{4z L_P R_Q}{R_N\nu(0)v_Fh}}
\end{equation}
where  $R_Q=h/4e^2$ is the quantum resistance.$R_N$ is related to the normal state array resistance $r_N$ by $r_N=\frac{M_x}{(M_y+1)(M_z+1)}R_N$ where the $x$ axis is along the array, $y,z$ are across it and $M_i$ is the number of layers in the $i$ direction\cite{Fazio2001}.  

Including this factor into the Gutzwiller trace formula (\ref{GF}), making the semi-classical approximation and evaluating for spherical grains, one finds the oscillating term in the density of states for the open system is given by\cite{Brack1997}, 
\begin{widetext}
\begin{equation}\label{DOS3}
\delta g(\epsilon)=\frac{3}{2}\sqrt{\frac{\pi}{kR}}\sum_{w=1}^\infty\sum_{v=2w}^\infty(-1)^w\sin(2\theta_{v,w})\sqrt{\frac{\sin{\theta_{vw}}}{v}}\sin{\Theta_{vw}}e^{-\frac{4 z L^{v,w}_p R_Q}{R_N \nu(0) v_Fh }}-\frac{3}{4}\frac{1}{k R}\sum_{w=1}^\infty\frac{1}{w}\sin(4wkR)e^{-\frac{4 z L^w_p R_Q}{R_N \nu(0) v_Fh }}
\end{equation}
\end{widetext}
where we have introduced the periodic orbit variables: $v$, the vertex number, $w$,  the winding number and the orbit length $L_P^{v,w}=2vR\sin{\theta_{v,w}}$, $\theta_{v,w}=\pi w/v$ and $\Theta_{v,w}=k L_P^{v,w}-3v\frac{\pi}{2}+\frac{3\pi}{4}$. The second sum concerns diameter orbits with orbit length $L_P^{w}=4wR$. The increase of the tunneling probability has the effect of suppressing oscillations eventually resulting in a smooth density of states. Henceforth we will use the compact notation,
\beq\label{compact}
\delta g(\epsilon)=\sum_{v,w}\tilde{g}_{v,w}^{(\frac{1}{2})}+\sum_w\tilde{g}_w^{(1)}
\eeq
where $\tilde{g}_{v,w}^{(\frac{1}{2})}$ and  $\tilde{g}_w^{(1)}$ correspond to the first and second terms in Eq. (\ref{compact}) respectively. The validity of this expression is restricted to the tight binding limit but it still provides a good description of tunneling in realistic metal oxide barriers\cite{strehlow1973}.

\subsection{The gap and critical temperature in coupled grains}

In this section we determine closed expressions for the gap and critical temperature in the coupled grain. First we determine the zero temperature gap by solving the gap equation,
\beq
1=\frac{\lambda }{2}\int_{-\epsilon_D}^{\epsilon_D}\frac{1+\bar{g}(0)+\delta g(\epsilon')}{\sqrt{\epsilon'^2+\Delta(R,0)^2}}d\epsilon'
\eeq
with the ansatz $\Delta(R,0)=\Delta_0(0)(1+f^{(\frac{1}{2})}+f^{(1)})$, expanding about the Fermi energy and solving by order in $(k_FL)^{-\frac{1}{2}}$ we find,
\beq
\begin{split}
f^{(\frac{1}{2})}=&\sum_{v,w}\tilde{g}_{v,w}^{(\frac{1}{2})}(0)K_0(\frac{L_P^{vw}}{\xi})\\
f^{(1)}=&\frac{\bar{g}^{(1)}(0)}{\lambda}+f^{(\frac{1}{2})}\left(\frac{f^{(\frac{1}{2})}}{2}-\sum_{v,w}\tilde{g}_{v,w}^{(\frac{1}{2})}(0)\frac{L_P^{vw}}{\xi}K_1(\frac{L_P^{vw}}{\xi})\right)\\ &+\sum_w\tilde{g}_w^{(1)}(0)K_0(\frac{L_P^{w}}{\xi})
\end{split}
\eeq
where $K_i(x)$is the $i^{th}$ order modified Bessel function of the second kind.

The critical temperature of the grain is found by solving the gap equation, Eq. (\ref{GapEqn}), at $\Delta\rightarrow 0$ and expanding about the Fermi energy, which gives
\begin{widetext}
\begin{equation}
k_BT_c(R)=\frac{2\epsilon_De^\gamma}{\pi}\Exp\left[-\frac{1}{\lambda(\bar g(0)+\sum_{v,w}\tilde{g}_{v,w}^{(\frac{1}{2})}\omega(L_P^{vw},T)+\sum_w\tilde{g}_w^{(1)}\omega(L_P^{w},T))}\right]
\end{equation}
\end{widetext}
where $\gamma$ is the Euler-Mascheroni constant and the weight function is given by,
\begin{equation}
\omega(L_P,T)=\frac{1}{\ln(\frac{2e^\gamma\beta\epsilon_D}{\pi})}\int_0^{\epsilon_D}\frac{1}{\epsilon}\cos(\frac{\epsilon}{2\epsilon_F} k_FL_P)\tanh(\frac{\beta\epsilon}{2})d\epsilon
\end{equation}
Hence we can express the gap close to the critical temperature by modifying the the BCS expression,
\begin{equation}
\Delta(R,T)\approx 1.74\Delta(R,0)\left(1-\frac{T}{T_c(R)}\right)^\frac{1}{2}
\end{equation}
The inclusion of the Bessel functions and weight function $\omega(L_P,T)$ is to suppress the oscillating term for long periodic orbits.
\subsection{The weak coupling limit}
We place a limit on the validity of the weak coupling approximation by finding $|T|^2$ using the Bardeen Transfer Hamiltonian formalism then combining with $R_N=(4\pi e^2|T|^2\nu(0)^2/\hbar)^{-1}$ one finds,
\beq
R_N=\frac{648 e^{2\kappa a}}{(\kappa R)^2(k_FR)^2} R_Q
\eeq
where we have assumed that the combined area of a grains junctions for a FCC lattice covers $\frac{1}{3}$ of the grains surface.  This puts the weak coupling $\kappa a \sim 1$ limit for our system at $R_N\sim 4\Omega$. Note also that the limit for weak coupling is well below the typical experimental parameters for a  metal oxide barrier $\kappa\sim 1.2\AA^{-1}$\cite{strehlow1973},  $a\sim 4\AA$, making the weak coupling limit ideal for studying the current problem.

\subsection{Phase-fluctuation driven transition}
 We will consider the usual action for such an array\cite{Chakravarty1987},
\begin{widetext}
\beq\label{action}
S=\frac{1}{2}\int_0^\beta d\tau\sum_i\frac{\dot{\phi_i^2}}{E_Q}-\frac{1}{2}\sum_{\langle ij\rangle}\int_0^\beta d\tau J_{ij}\cos(2(\phi_i(\tau)-\phi_j(\tau)))+2\sum_{\langle ij\rangle}\int_0^\beta d\tau\int_0^\beta  d \tau' G_{ij}(\tau-\tau')\sin^2(\frac{1}{4}(\delta\phi_{ij}(\tau)-\delta\phi_{ij}(\tau')))
\eeq
\end{widetext}
where $\phi_i$ is the phase in grain $i$ and $\delta\phi_{ij}=\phi_i-\phi_j$. The first term is the charging energy, the second is the Josephson coupling and the final term accounts for quasi-particle tunneling. The charging energy is $E_Q=\frac{(2e)^2}{C}$.
This action has been studied at length in the literature. Here we follow the solution of Panyukov and Zaikin\cite{Panyukov1987} and map the action onto a Ginzberg-Landau free energy. To solve the problem in a non homogeneous array we start by moving to a mean-field theory in the gap to remove the position dependence of $J_{ij}, G_{ij}$. The problem is then identical to that treated in\cite{Panyukov1987}  with these modified definitions.\\
The Josephson energy is defined in general by\cite{Ambegaokar1963},
\begin{equation}
J_{ij}=\frac{\Delta_i\Delta_j}{\beta}\frac{R_Q}{R_N}\sum_{l=-\infty}^{\infty}\frac{1}{\sqrt{((\frac{\pi(2l+1)}{\beta})^2+\Delta_i^2)((\frac{\pi(2l+1)}{\beta})^2+\Delta_j^2)}}
\end{equation}
We re-parameterize this expression using the mean gap across the junction and the separation of the gaps across the junction; respectively given by $\bar\Delta_{ij}=\frac{\Delta_i+\Delta_j}{2}, \Delta_{ij}'=\frac{|\Delta_i-\Delta_j|}{2}$. We then  make an expansion about $\Delta'=0$ to find,
\begin{widetext}
\begin{equation}\label{J}
J_{ij}=\frac{\bar\Delta_{ij} R_Q}{2R_N}\tanh(\frac{\beta\bar\Delta_{ij}}{2})-(\frac{3}{\bar\Delta_{ij}}\tanh(\frac{\beta\bar\Delta_{ij}}{2})+\frac{\beta}{2}\sech^2(\frac{\beta\bar\Delta_{ij}}{2})+\frac{\beta^2\bar\Delta_{ij}}{2}\sech^2(\frac{\beta\bar\Delta_{ij}}{2})\tanh(\frac{\beta\bar\Delta_{ij}}{2}))\frac{\Delta_{ij}'^2R_Q}{8R_N}
\end{equation}
\end{widetext}
Similarly we can consider the quasiparticle term. In all cases of interest $k_BT\ll\mu$ so we may work in a zero temperature approximation where\cite{Eckern1984},
\beq
G_{ij}(\tau)=\frac{\hbar}{2\pi e^2R_N}\frac{\Delta_i\Delta_j}{\hbar^2}K_1\left(\frac{\Delta_i|\tau|}{\hbar}\right)K_1\left(\frac{\Delta_j|\tau|}{\hbar}\right)
\eeq
 Noting the localized nature of $G(\tau)$ and assuming phase varies slowly on scale $\hbar/\Delta$ we can perform a gradient expansion, in $\tau$, to express the dissipative part of the action as,
\begin{widetext}
\beq
\frac{S_G}{\hbar}=\sum_{\langle ij\rangle}\frac{\Delta_j\hbar^2}{16e^2R_N}\left(\frac{(\Delta_i^2+\Delta_j^2)\elE(\sqrt{1-\frac{\Delta_i^2}{\Delta_j^2}})-2\Delta_i^2\elK(\sqrt{1-\frac{\Delta_i^2}{\Delta_j^2}})}{(\Delta_i^2-\Delta_j^2)^2}\right)\int_0^{\beta\hbar}d\tau(\frac{\partial}{\partial\tau}\delta\phi_{ij}(\tau))^2
\eeq
\end{widetext}
where $\elK,\elE$ are complete elliptic integrals of the first and second kind respectively and we have assumed without loss of generality that $\Delta_i<\Delta_j$. $S_G$ is a monotonic function of $\Delta'$ varying $S_G$ by a prefactor from $\frac{3\pi}{8}\rightarrow 1$
as $\Delta'$ goes from $0\rightarrow\bar{\Delta}$ respectively. As this modification is very small in general we will make the approximation that the prefactor is $\frac{3\pi}{8}$ and hence,
\beq
\frac{S_G}{\hbar}=\sum_{\langle ij\rangle}\frac{3\pi\hbar^2}{256e^2\bar\Delta_{ij} R_N}\int_0^{\beta\hbar}d\tau(\frac{\partial}{\partial\tau}\delta\phi_{ij}(\tau))^2
\eeq
in all cases. This simplification allows us to treat the superconducting grains and the normal grains around the superconducting cluster identically, effectively removing any need to consider the actual arrangement of the grains.
We can now move to a mean-field theory where we approximate the disordered infinite cluster by a regular array with gaps given by the mean properties across all the junctions. For the Josephson term we reduce the nearest neighbor sum to the superconducting nearest neighbors however for the charging and dissipation term we continue to use all nearest neighbors as for these terms the behavior is not affected by the phase. We move to a mean field theory by dropping the indicies $i,j$ and replacing these properties with the following mean-value definitions taken over the superconducting cluster, 
\beq
\bar\Delta_{ij}\rightarrow\bar\Delta=\frac{1}{p^2}\int_0^\infty\int_0^\infty\frac{\Delta(R)+\Delta(R')}{2}P_{sc}(R)P_{sc}(R')dRdR'
\eeq
\beq
 \Delta_{ij}'\rightarrow\Delta'=\frac{1}{p^2}\int_0^\infty\int_0^\infty\frac{|\Delta(R)-\Delta(R')|}{2}P_{sc}(R)P_{sc}(R')dRdR'
\eeq
where p is the fraction of all the grains which are superconducting, $p=\int_0^\infty P_{sc}(R)dR$. This is a good approximation as close to the transition the distribution of $\bar\Delta_{ij}$ and $\Delta'_{ij}$ in the array will be narrow and sharply peaked. With these new definitions the Josephson tunneling energy, Eq. \ref{J}, becomes,
\begin{widetext}
\begin{equation}
\begin{split}
J=&\frac{\bar\Delta R_Q}{2R_N}\tanh(\frac{\beta\bar\Delta}{2})-\Lambda\\
\Lambda=&(\frac{3}{\bar\Delta}\tanh(\frac{\beta\bar\Delta}{2})+\frac{\beta}{2}\sech^2(\frac{\beta\bar\Delta}{2})+\frac{\beta^2\bar\Delta}{2}\sech^2(\frac{\beta\bar\Delta}{2})\tanh(\frac{\beta\bar\Delta}{2}))\frac{\Delta'^2R_Q}{8R_N}
\end{split}
\end{equation}
\end{widetext}

 We can now re-express the action Eq. (\ref{action}) by making a Fourier transform as,
\beq
\begin{split}
\frac{S}{\hbar}=&\frac{1}{2}\frac{\hbar^2}{k_BT}\sum_{\bf{q}\;\omega}\left(\frac{\omega^2}{E_Q}+\frac{\omega^2}{E_Q^*}\sum_{\boldsymbol{\alpha}}\lambda_{\boldsymbol{\alpha}}(q)\right)|\phi({\bf q,\omega})|^2\\ &-\frac{1}{2}\sum_{\langle ij\rangle_{sc}}\int_0^\beta d\tau J\cos(2(\phi_i(\tau)-\phi_j(\tau)))
 \end{split}
\eeq\\
\\
\\
where $\omega=\frac{2\pi n k_BT}{\hbar},\;n=0,\pm1,\pm2\ldots$ and $\lambda_{\boldsymbol{\alpha}}({\bf q})=1-e^{i{\bf q}\cdot \boldsymbol{\alpha}}$, $\boldsymbol{\alpha}$ are the lattice vectors and $E_Q^*=\frac{124e^2\bar\Delta R_N}{3\pi\hbar}$. Note we have modified the sum in the Josephson coupling term to just the superconducting nearest neighbors however all other terms maintain their sum over all neighboring grains. Making a Hubbard-Stratonovich transformation in the Josephson term and integrating out $\phi_i(\tau)$ we find in the limit ${\bf q}\rightarrow 0, \omega\rightarrow0$,
\begin{widetext}
\beq
\frac{S}{\hbar}=\frac{\bar zJ}{2k_BT}\left(1-\frac{\bar z J}{2}\int_0^{\beta\hbar}\frac{d\tau}{\hbar} X_{ii}(\tau)\right)|\psi({\bf q}=0,\omega=0)|^2+\frac{\zeta}{4}|\psi({\bf q}=0, \omega=0)|^4
\eeq
\end{widetext}
where $\psi$ is the Hubbard-Stratonovich field, $X_{ii}$ is the correlation function,
\beq
\ln{X_{ii}(\tau)}=-\frac{1}{2}\langle(\phi_i(\tau)-\phi_i(\tau'))^2\rangle=\frac{\tilde E_Q\tau}{2\hbar},
\eeq
 $\zeta$ is a numerical constant, $\tilde E_Q=(\frac{1}{E_Q}+\frac{\eta}{E_Q^*})^{-1}$, $\eta$ is an integration constant from integrating out $\bf q$ and $\bar z$ is the mean number of superconducting neighbor grains in the percolating cluster, $\bar z=z p$. The critical temperature is found by solving,
\begin{equation}\label{CritCond}
1=\frac{\tilde E_Q}{ \bar z J}+e^{-\frac{\tilde E_Q}{2 k_BT}}
\end{equation}

\subsection{Specific Heat}
We calculate the specific heat of a superconductor from the electronic entropy,
\begin{equation}\label{entropy1}
S=-2k_B\sum_k((1-f_k)\ln(1-f_k)+f_k\ln(f_k))
\end{equation}
where $f_k=(1+e^{\beta E_K})^{-1}$ is the dirac-distribution. For a superdoncudtor $E_k=\sqrt{\epsilon^2_k+\Delta(T)^2}$. 
We can calculate the entropy in the region of the condensate by restricting the sum in \ref{entropy1} to the region within the Debye energy of the Fermi surface. 
The electronic specific heat is defined by,
\begin{equation}
\begin{split}\label{SH}
&C_\text{es}=-\beta\frac{dS}{d\beta}\\
&C_\text{es}=2\beta k_B\int_{-\epsilon_D}^{\epsilon_D}-\frac{\partial f(\epsilon)}{\partial E(\epsilon)}\left(\epsilon^2+\Delta^2\left(1-\frac{d\ln\Delta}{d\ln T}\right)\right)\nu(\epsilon)d\,\epsilon
\end{split}
\end{equation}
In the limit $\Delta\to0$ we recover the usual normal metal electronic specific heat $C_\text{en}\propto T$. We include the size dependent density of states using Eqs. (\ref{DOS1}),  (\ref{DOS2}) and (\ref{DOS3}). Taking $\nu(0)$ to be the total, volume independent spectral density of the grain Eq. (\ref{SH}) computes the volume independent specific heat of each grain.

To calculate the specific heat of the array as a function of volume we integrate over all grain sizes and divide by the total volume of the grains,
\begin{equation}
C^\text{Array}_\text{es}=\frac{\int^{\infty}_{-\infty}C_\text{es}P(R)d\,R}{\int^{\infty}_{-\infty}V(R)P(R)d\,R}
\end{equation}
where $V(R)=\frac{4}{3}\pi R^3$.
Note that for an inhomogeneous array grains go from superconductors to normal metals progressively and thus the usual sharp peak observed for a bulk superconductor should become smoothed. This is a hallmark of an inhomogeneous transition.
\bibliography{library}
\end{document}